\documentclass[preprint,unsortedaddress,showpacs,preprintnumbers,amsmath,amssymb,nofootinbib]{revtex4-1}
\usepackage{graphicx}
\usepackage{subfigure}
\usepackage{dcolumn}
\usepackage{bm}
\usepackage[T1]{fontenc}
\usepackage[utf8]{inputenc}
\usepackage[english]{babel}
\usepackage{blindtext}

\def \be {\begin{equation}}
\def \ee {\end{equation}}
\def \bea{\begin{eqnarray}}
\def \eea{\end{eqnarray}}

\begin{document}
\title{Orbit analysis of a geostationary \\gravitational wave
  interferometer detector array}
\author{Massimo Tinto} 
\email{massimo.tinto@jpl.nasa.gov}
\affiliation{Jet Propulsion Laboratory, California
  Institute of Technology,\\ Pasadena, CA 91109}
\author{Jose C. N. de Araujo}
\email{jcarlos.dearaujo@inpe.br}
\affiliation{Instituto Nacional de Pesquisas Espaciais,\\
S. J. Campos, SP, Brazil}
\author{Helio K. Kuga}
\email{hkk@dem.inpe.br}
\affiliation{Instituto Nacional de Pesquisas Espaciais,\\ 
S. J. Campos, SP, Brazil}
\author{M\'arcio E. S. Alves}
\email{alvesmes@gmail.com}
\affiliation{Instituto de F\'{i}sica e Qu\'{i}mica, Universidade Federal de
Itajub\'{a},\\
Itajub\'{a}, MG, Brazil}
\author{Odylio D. Aguiar}
\email{odylio.aguiar@inpe.br}
\affiliation{Instituto Nacional de Pesquisas Espaciais,\\
S. J. Campos, SP, Brazil}

\date{\today}
\begin{abstract}
 
  We analyze the trajectories of three geostationary satellites
  forming the GEOstationary GRAvitational Wave Interferometer
  (GEOGRAWI)~\cite{tinto}, a space-based laser interferometer mission
  aiming to detect and study gravitational radiation in the ($10^{-4}
  - 10$) Hz band.
  
  The combined effects of the gravity fields of the Earth, the Sun and
  the Moon make the three satellites deviate from their nominally
  stationary, equatorial and equilateral configuration. Since changes
  in the satellites relative distances and orientations could
  negatively affect the precision of the laser heterodyne
  measurements, we have derived the time-dependence of the
  inter-satellite distances and velocities, the variations of the
  polar angles made by the constellation's three arms with respect to
  a chosen reference frame, and the time changes of the triangle's
  enclosed angles.  We find that, during the time between two
  consecutive station-keeping maneuvers (about two weeks), the
  relative variations of the inter-satellite distances do not exceed a
  value of $0.05$ percent, while the relative velocities between pairs
  of satellites remain smaller than about $0.7 \ {\rm m/s}$.  In
  addition, we find the angles made by the arms of the triangle with
  the equatorial plane to be periodic functions of time whose
  amplitudes grow linearly with time; the maximum variations
  experienced by these angles as well as by those within the triangle
  remain smaller than $3$ arc-minutes, while the East-West angular
  variations of the three arms remain smaller than about $15$
  arc-minutes during the two-weeks period. The relatively small
  variations of these orbit parameters result into a set of system
  functional and performance requirements that are less stringent than
  those characterizing an interplanetary mission.

\end{abstract}


\maketitle

\section{Introduction}
\label{intro}
Gravitational waves (GWs), predicted by Einstein's theory of general
relativity, are disturbances of space-time propagating at the speed of
light. Because of their extremely small cross-sections, GWs carry
information about regions of the Universe that would be otherwise
unobtainable through the electromagnetic spectrum.  Once detected, GWs
will allow us to open a new observational window on the Universe, and
perform a unique test of general relativity~\cite{Thorne}.

Since the first pioneering experiments by Joseph Weber in the early
sixties~\cite{Weber}, several experimental groups around the world
have been attempting to detect GWs. The coming on-line of the Advanced
LIGO and Advanced Virgo detectors~\cite{LIGO,VIRGO}, however, is
likely to break this sequence of experimental drawbacks and result
into the first detection before the end of this decade.

Contrary to ground-based detectors, which are sensitive to
gravitational waves in a band from about a few tens of Hz to a few
kilohertz, space-based interferometers are expected to access a much
lower frequency region (from a few tenths of millihertz to about a few
tens of Hz) where GW signals are expected to be larger in number and
characterized by larger amplitudes. The most notable example of a
space interferometer, which for several decades has been jointly
studied in the United States of America and in Europe, is the Laser
Interferometer Space Antenna (LISA) mission~\cite{LISA}.  By relying
on coherent laser beams exchanged between three remote spacecraft
forming a giant (almost) equilateral triangle of arm-length equal to $5
\times 10^6$ km, LISA aimed to detect and study cosmic gravitational
waves in the $10^{-4} - 1$ Hz band.

Although over the years only a few space-based detector designs have
been considered as alternatives to the LISA
mission~\cite{ASTROD,DECIGO,Hiscock} starting in 2011 other mission
concepts have appeared in the literature~\cite{PCOS} in conjunction
with the ending of the NASA/ESA partnership for flying LISA.  Their
goals were to meet (at a lower cost) most (if not all) the LISA
scientific objectives (highlighted in the 2010 Astrophysics Decadal
Survey {\it New Worlds, New Horizons (NWNH)}~\cite{NWNH}).

GEOGRAWI was one of the alternative concepts to the LISA mission
submitted in response to NASA's Request for Information \#
NNH11ZDA019L~\cite{RFI}. It entails three spacecraft in geostationary
orbit, forming an equilateral triangle of arm-length approximately
equal to $73,000$ km.  Like LISA, it has three identical spacecraft
exchanging coherent laser beams but, by being in a geostationary
orbit, it achieves its best sensitivity in a frequency band ($3 \times
10^{-2} - 1$ Hz)~\cite{tinto} that is complementary to those of LISA
and ground detectors. The astrophysical sources that GEOGRAWI is
expected to observe within its operational frequency band include
extra-galactic massive and super-massive black-hole coalescing
binaries, the resolved galactic binaries and extra-galactic coalescing
binary systems containing white dwarfs and neutron stars, a stochastic
background of astrophysical and cosmological origin, and possibly more
exotic sources such as cosmic strings. GEOGRAWI will be able to test
Einstein's theory of relativity by comparing the waveforms detected
against those predicted by alternative relativistic theories of
gravity, and also by measuring the number of independent polarizations
of the detected gravitational wave signals~\cite{TintoAlves2010}.

In order to compensate for the gravitational perturbations exerted by
the Sun, the Moon, and the gravity field of the Earth (which would
result into a long-term orbital drift), a geostationary satellite must
perform regular operations of ``station-keeping''~\cite{Soop}. This
entails firing the onboard thrusters to keep the satellite at its
required location. In the case of GEOGRAWI the station-keeping
maneuvers performed by its three satellites are particularly important
as they offset the excessive variations of the inter-spacecraft
relative orientations and velocities and maintain the constellation in
a stable configuration. If the spacecraft would be left to drift apart
for periods longer than the station-keeping duty cycle, the quality of
the laser heterodyne measurements performed by the constellation would
degrade, resulting into a degradation of the science objectives of
GEOGRAWI~\cite{tinto}.

Our paper analyzes the time evolution of the GEOGRAWI constellation
during the time between two station-keeping maneuvers, and it is
organized as follows. In section~\ref{orbit} we derive the trajectory
of each spacecraft (as a function of time) by numerically solving the
Newtonian equations of motion of a ``point-particle'' in the
gravitational potentials of the Earth, the Moon and the Sun. After
noticing that the effects of the solar radiation pressure on each
spacecraft are not included into the equations of motion because they
are compensated for by the spacecraft drag-free system, in
section~\ref{disvel} we estimate the magnitude of the variations of
the inter-spacecraft distances, velocities, and relative angular
orientations. We find that, during the time between two consecutive
station-keeping maneuvers (about two weeks), the relative variations
of the inter-satellite distances do not exceed a value of $0.05$
percent, while the relative velocities between pairs of satellites
remain smaller than about $0.7 \ m/s$.  In addition, we find the
angles made by the arms of the triangle with the equatorial plane to
be periodic functions of time whose
amplitudes grow linearly with time; the maximum variations experienced
by these angles as well as by those within the triangle remain smaller
than $3$ arc-minutes, while the East-West angular variations of the
three arms remain smaller than about $15$ arc-minutes during a
two-weeks period.  These relatively small variations of the orbit
parameters result into a set of system performance and functional
requirements that are less stringent than those characterizing an
interplanetary mission.

\section{Orbit dynamics}
\label{orbit}
GEOGRAWI measures relative frequency changes experienced by coherent
laser beams exchanged by its three pairs of spacecraft.  As the laser
beams are received, they are made to interfere with the outgoing laser
light. These heterodyne measurements are each down-converted with the
use of an onboard oscillator, then digitized and numerically combined
in order to cancel the lasers frequency
fluctuations~\cite{TD2005}. The spacecraft are made to follow an orbit
determined by the gravitational forces on a (spherical) test mass
(located onboard each spacecraft) due to the Earth, the Moon and the
Sun, and are therefore drag-free. Since the relative distances between
spacecraft are not constant, the heterodyne measurements will display
a resulting Doppler shifts, which is then removed from the data by
using an on-board clock. The magnitude of the frequency fluctuations
introduced by the clocks into the heterodyne measurements depends
linearly on the noises from the clocks themselves and the
inter-spacecraft relative velocities.  Space-qualified, state of the
art clocks are oven-stabilized crystals characterized by an Allan
deviation of ${\sigma_A} \approx 10^{-13}$ for averaging times of $1 -
1000$ s, covering most of the frequency band of interest to
GEOGRAWI. The corresponding power spectral density of the relative
frequency fluctuations, $S_y(f)$, associated with this
``flicker-noise'' at the Fourier frequency $f$ is given by the
following expression~\cite{Barnes}
\begin{equation}
S_y (f) = { \frac{{\sigma_A}^2}{2 ln 2}} { \frac{\nu_D^2}{\nu_0^2}}
\ f^{-1} \ {\rm Hz}^{-1} \ ,
\label{SN}
\end{equation}
where we have denoted with $\nu_D$ the frequency change
induced by the Doppler effect on the one-way heterodyne measurement
and with $\nu_0$ the nominal laser frequency. Since $\nu_D$ is equal
to $\nu_D = \nu_0 \ v/c$, with $v$ being the two spacecraft relative
velocity and $c$ the speed of light, we can rewrite Eq.(\ref{SN}) in
the following form
\begin{equation}
S_y (f) = { \frac{{\sigma_A}^2}{2 ln 2}} { \frac{v^2}{c^2}}
\ f^{-1} \ {\rm Hz}^{-1} \ .
\label{SNnew}
\end{equation}
If we take a spacecraft relative velocity of $0.7 \ {\rm m/s}$ (which
we derive in the following subsection~\ref{disvel}), a Fourier
frequency $f = 10^{-3} \ {\rm Hz}$, and the frequency of the laser to
be equal to $ \nu_0 = 3.0 \ \times 10^{14} \ {\rm Hz}$, we find a $S_y
(10^{-3}) = 3.9 \times 10^{-41} \ {\rm Hz}^{-1}$. To put this number in
perspective, in reference~\cite{tinto} we showed that at this
frequency the GEOGRAWI noise level goal (determined by our estimate of
the remaining noises) was equal to about $3.7 \ \times 10^{-46}$,
roughly five orders of magnitude smaller than the above estimated
noise due to the clock. Recent developments in the design of
space-qualified clocks~\cite{Prestage} indicate that oscillators
capable of an Allan deviation of a few parts in $10^{-15}$ or better
at 1000 s integration time will be available by the end of this
decade. Such a clock performance would still require the
implementation of the clock-noise calibration procedure~\cite{TEA02}
with an inter-spacecraft velocity of $0.7 \ {\rm m/s}$ since the
resulting clock noise spectrum would still be about a factor of $10$
or so above the GEOGRAWI noises. It should be said, however, that the
resulting set of performance requirements levied on the onboard
subsystems involved in the clock calibration procedure will be less
stringent than those characterizing an interplanetary
mission~\cite{LISA} as its inter-spacecraft velocities are
approximately $30$ times larger than those of GEOGRAWI.

Besides the magnitude of the inter-spacecraft relative velocities, the
spacecraft ability of properly pointing to each other is obviously
important. In the case of LISA~\cite{LISA}, for instance, it was shown
that the trajectories of its three spacecraft resulted into a
variation of the angles enclosed by the constellation's triangle of
about $\pm 1^0$. This required the implementation of a mechanical
system for articulating the two optical telescopes onboard each
spacecraft in such a way to maintain the optical links across the
constellation.

In order to address the above questions in the contest of GEOGRAWI, in
what follows we first integrate the equations of motion for each of
the three GEOGRAWI spacecraft. At an arbitrarily chosen starting time
$t = 0$, we assume the spacecraft to be at rest with respect to a
right-handed, orthogonal coordinate system associated with a reference
frame that rotates jointly with the Earth. In it the $Z$-axis
coincides with the Earth's axis of rotation, the $X$-axis intersects
the Greenwich line in the equatorial plane, and the $Y$ axis is
orthogonal to it. Under the influence of the Earth, Moon, and Sun
gravitational fields, the three spacecraft move from their starting
locations corresponding to an equilateral triangle configuration.

Since the Earth is much closer to the spacecraft than the Moon and the
Sun, its gravitational potential has to be described
mathematically in such a way to reflect the Earth's mass non-spherical
and non-symmetrical distribution.  This uneven distribution of mass is
expressed by the so-called coefficients of spherical harmonics, which
enter into the multi-pole expansion of the Earth's gravitational
potential (discussed in the appendix). In order to have high
accuracy, and because it was also easy to do, our model of the Earth's
gravitational potential relied on the EGM2008, which was made to
include terms in the multi-pole expansion up to order/degree
2100~\cite{pavlis}. By relying on our numerical
integrator~\cite{kuga}, we derived the trajectories of the three
satellites, i.e. their location and velocity vectors as functions of
time. In what follows we provide a description of the kinematic
quantities of relevance to GEOGRAWI that we have been able to derive,
i.e. the time-dependence of the inter-satellite distances and
velocities, and their relative angular orientations. 

\subsection{Inter-satellite distances, velocities, and relative orientations}
\label{disvel}
Before describing our estimated relative changes of the three
arm-lengths, we should first remind ourselves that the gravity field
of the Earth is not invariant under rotation around the Earth's
rotation axis because of its non-spherical and non-symmetrical mass
distribution. This implies that the variation of the inter-spacecraft
distances will depend on the initial longitudinal configuration of the
constellation, opening the possibility for the existence of a specific
starting longitude of the equilateral triangle minimizing the maximum
of the inter-spacecraft velocities. Although there exist a specific
value of the longitude of the constellation resulting into a mini-max
value of the relative velocities, the resulting reduction in the
relative velocities is not significant enough to make it a mission
requirement.

\begin{figure}
\includegraphics[width=4.0in, angle=0]{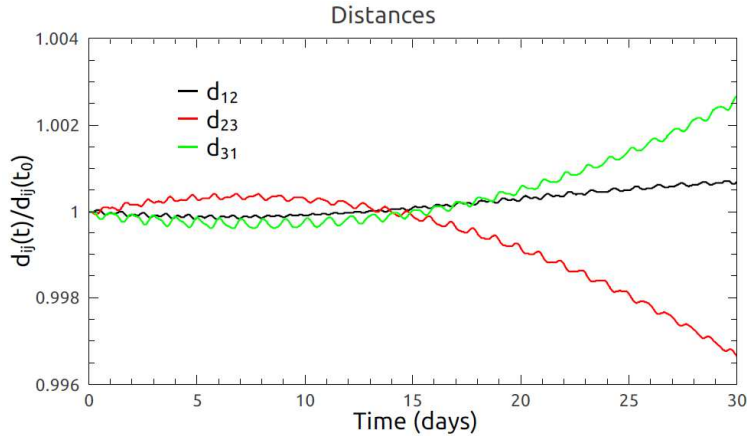}
\caption{Relative variation of the three inter-spacecraft distances
  estimated during a period of thirty days. The nominal starting
  configuration of GEOGRAWI is taken to be an equilateral
  triangle of arm-length $7.3 \times 10^4 \ {\rm km}$.}
\label{fig1}
\end{figure}

In figure (\ref{fig1}) we plot the time-dependence of the three
inter-spacecraft distances relative to the nominal arm-length of the
equilateral triangle at time $t = 0$ ($7.3 \times 10^4 \ {\rm
  km}$). The plot covers a period of thirty days during which no
station-keeping maneuvers have been accounted for. Since these
typically happen about once or twice per fortnight, we conclude that
during the first, say, fifteen days the maximum relative variations of
the three inter-spacecraft distances do not exceed a value of $0.05$
percent. Although such small variations of the relative distances,
together with the resulting arm-length inequalities, still require
GEOGRAWI to rely on Time-Delay Interferometry (TDI)~\cite{TD2005} for
canceling the laser frequency fluctuation, more stable coherent lasers
expected to become space-qualified before the end of this
decade~\cite{Yu} should allow us to operate GEOGRAWI like a
ground-based interferometer.

\begin{figure}
\includegraphics[width=4.0in, angle=0]{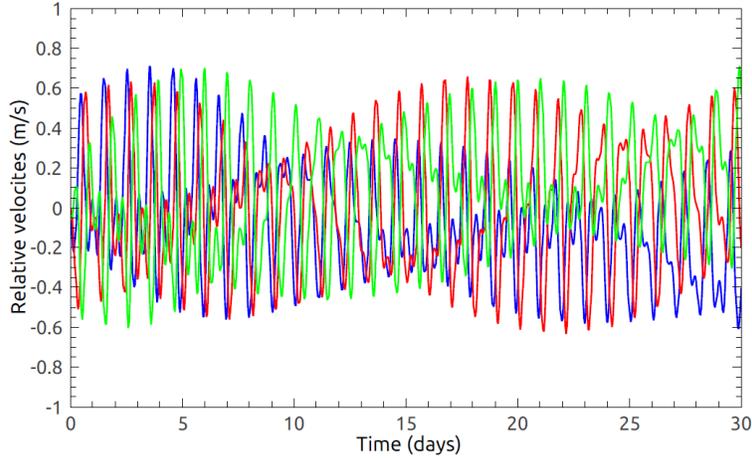}
\caption{Time-dependence of the inter-spacecraft velocities estimated
  during a period of thirty days. No station-keeping maneuvers have
  been accounted for during this time period.}
\label{fig2}
\end{figure}

In figure (\ref{fig2}) we now plot the time-dependence of the
inter-spacecraft relative velocities, again for a period of thirty
days. Note these three time-series are periodic functions of time with
period equal to one day and amplitudes that depend on time and do not
exceed a maximum value of about $0.7 \ {\rm m/s}$. Although this is
still large enough to require the use of the calibration procedure for
canceling the noise from current space-qualified clocks in the
GEOGRAWI interferometric measurements, it is still significantly
(about a factor of $30$) smaller than the values characterizing
interplanetary missions~\cite{LISA}, making the procedure for its
calibration easier to be implemented.

\begin{figure}
\includegraphics[width=4.0in, angle=0]{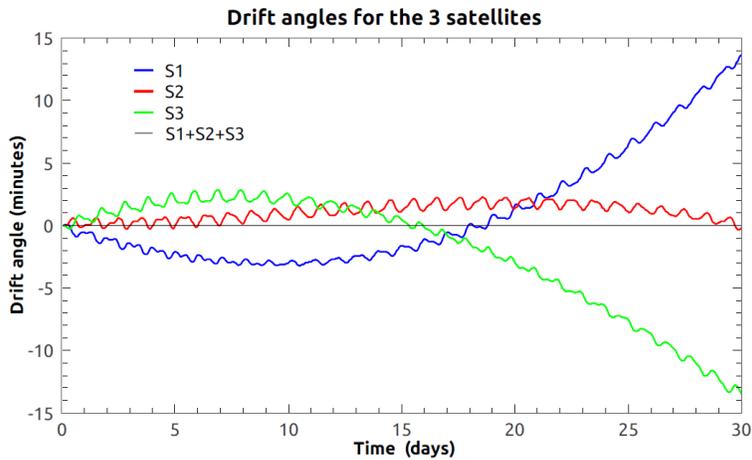}
\caption{Deviations of the angles enclosed by the constellation with
  respect to $60^0$, corresponding to the nominal equilateral
  configuration at time $t=0$.}
\label{fig3}
\end{figure}

In figure (\ref{fig3}) we now plot the variation of the angles
enclosed by the triangular constellation. The values shown correspond
to the differences between each angle's value at time $t$ and the
$60^0$ value at time $t=0$. During the first two weeks the enclosed
angles do not change much, remaining within the $\pm 3 \ {\rm
  arc-minute}$ range. To put this number in perspective, in the case
of the former LISA mission it was estimated a variation of its
enclosed angles of about $\pm 1^0$. In order for the LISA spacecraft
to track each other it was assessed that each spacecraft had to have an
articulation mechanism for varying the angle enclosed by the two
onboard optical telescopes. Because of a much smaller variation of its
enclosed angles, GEOGRAWI will not need to implement such articulation
mechanism.

\begin{figure}
\includegraphics[width=4.0in, angle=0]{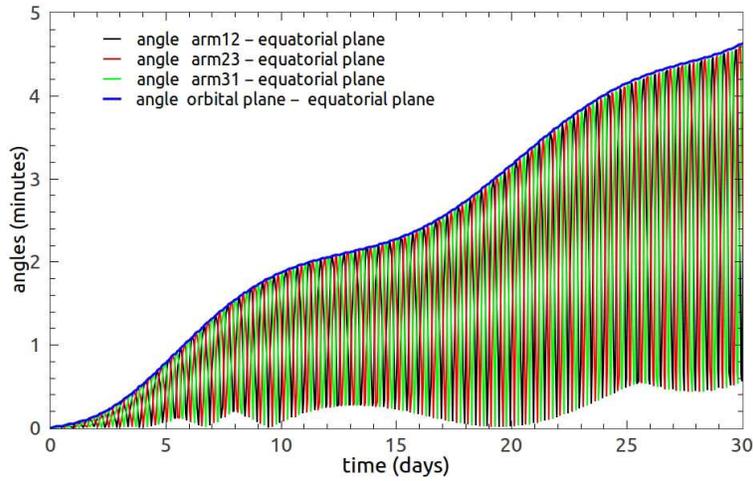}
\caption{Time-variation of the angles made by the three arms with
  respect to the equatorial plane. See text for a detailed description.}
\label{fig4}
\end{figure}

To complement the results shown in figure (\ref{fig3}) we have derived
the variation of the polar angles describing the orientation of each
arm of the interferometer. In figure (\ref{fig4}) we plot the (i)
time-dependence of the three polar angles describing the inclination
of the three arms relative to the equatorial plane, and (ii) the
inclination of the entire triangle relative to the equatorial
plane. The latter is found to be a monotonically growing function of
time, changing about $2 \ {\rm arc-minutes}$ during the first fifteen
days, while the three polar angles are periodic functions of time 
and monotonically increasing amplitudes. 

\begin{figure}
\includegraphics[width=4.0in, angle=0]{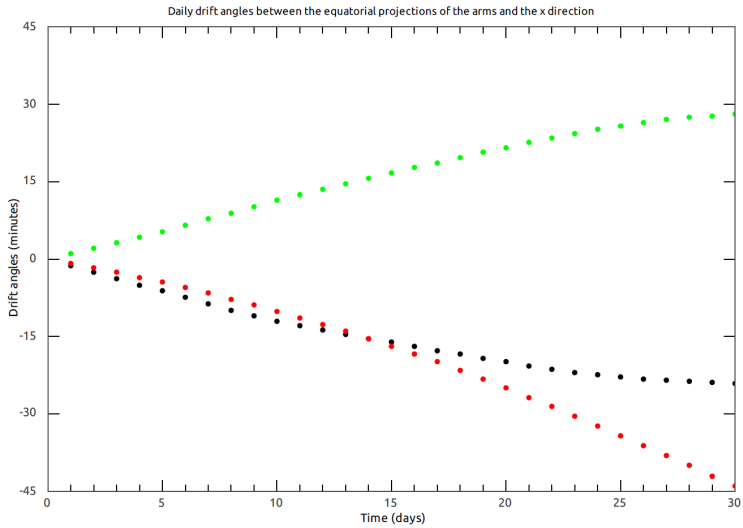}
\caption{Variation of the angles made by the projections of the three
  arms over the equatorial plane with an arbitrarily chosen $X-$ axis.}
\label{fig5}
\end{figure}

Finally in figure (\ref{fig5}) we plot the variation of the remaining
three polar angles made by the projections of the three arms on the
equatorial plane with respect to the $X-$axis. In this case the
variations are larger than those shown in figure (\ref{fig4}), but do
not exceed a maximum value of $15 \ {\rm arc-minutes}$ during a period
of two weeks.

\section{Conclusions}
\label{Conclusions}
In this article we have analyzed the trajectories of three
geostationary satellites forming the constellation of a laser
interferometer gravitational wave detector. We have found that, during
the time between two consecutive station-keeping maneuvers (about two
weeks), the relative variations of the inter-satellite distances do
not exceed a value of $0.05$ percent, while the relative velocities
between pairs of satellites remain smaller than about $0.7 \
m/s$. Since it is likely that future space-qualified clocks will be
characterized by a frequency stability that is more than two orders of
magnitude better than what currently available, such small relative
velocities might imply no need for implementing the clock-noise
calibration procedure. This would result into a significant
simplification of the hardware architecture that makes the clock-noise
calibration procedure possible.

In addition, we found the angles made by the arms of the triangle with
the equatorial plane to be periodic functions of time 
whose amplitudes grow linearly with time; the
maximum variations experienced by these angles as well as by those
within the triangle remain smaller than $3$ arc-minutes, while the
East-West angular variations of the three arms remain smaller than
about $15$ arc-minutes during a two-weeks period.  These relatively
small variations of the orbit parameters result into a set of system
functional and performance requirements that are less stringent than
those characterizing an interplanetary mission.
\section*{Acknowledgments}

MT acknowledges financial support provided by the Jet Propulsion
Laboratory Research \& Technology Development program. JCNA thanks
FAPESP and CNPq for partial financial support, while MESA acknowledges
financial support from FAPEMIG (Grant APQ-00140-12). For MT, this
research was performed at the Jet Propulsion Laboratory, California
Institute of Technology, under contract with the National Aeronautics
and Space Administration.

\appendix
\section{Orbital Gravitational Effects Computation}
\label{geopot}
A material point (body) subject to attraction by the non-central
gravitational field of the Earth suffers disturbances due to non-spherical
and non-symmetrical distribution of its mass. This uneven distribution
of mass is expressed by the so-called coefficients of spherical
harmonics, and the gravitational potential, $V$, of a body relative to the Earth is
given by the following general multi-poles expansion

\begin{equation}
V = \frac{GM_{e}}{r} \sum_{n=0}^{\infty} \sum_{m=0}^{M}
\left(\frac{a}{r} \right)^{n}
\left[\bar{C}_{nm}\cos (m\lambda) + \bar{S}_{nm}\sin (m\lambda) \right]
\bar{P}_{nm}(\sin \Psi) \ ,
\end{equation}

\par\noindent 
where $G$ is the universal gravitational constant, $M_{e}$ is the
Earth mass, $M$ is the truncation index, $r$ is the distance to the
body from the center of the Earth, $a$ is the Earth equatorial radius,
$\lambda$ is the longitude of the body, $\Psi$ is the geocentric latitude
of the body, $P_{nm}$ are the fully normalized Legendre polynomials of
order $n$ and degree $m$, and $\bar{C}_{nm}$ , $\bar{S}_{nm}$ are the
fully normalized spherical harmonics coefficients.

Older models of harmonics coefficients did not need analysis or
optimization for the numerical computation of the geopotential
model. The coefficients were in general of low order and
degree. Nowadays the gravitational models easily start from
order/degree 360 (like EGM96 model~\cite{lemoine}) and go up to more
than order/degree 2100 (like EGM2008 (Pavlis et al., 2008,
\cite{pavlis})). These computations require the calculation of the
Legendre polynomials, which should be recursively evaluated for high
order and degree.  Herein we used the standard-forward-column
implementation proposed by Holmes and Featherstone~\cite{holmes},
which is believed to be numerically superior and was described in Kuga
and Carrara~\cite{kuga} where its computation performance for Earth
orbits was verified.

To implement the algorithm it is convenient to reverse the order of
computation of the summation, where the outer loop in $m$ is first
computed. Let us rewrite the geopotential summation as:

\begin{eqnarray} \nonumber
V = \frac{GM}{r} +  \frac{GM}{r}\sum_{m=0}^{M} &&
\left[\cos(m\lambda)
\sum_{n=\mu}^{M} \left(\frac{a}{r} \right)^{n} \bar{C}_{nm} \bar{P}_{nm}(\Theta)+ \right.\\
&& \left. + \sin (m\lambda)
\sum_{n=\mu}^{M} \left(\frac{a}{r} \right)^{n} \bar{S}_{nm}
\bar{P}_{nm}(\Theta)\right],
\label{V}
\end{eqnarray}

\par\noindent 
where $ 0^{o} < \Theta < 180^{o}$ is now the co-latitude. It is
convenient to define the inner terms in equation (\ref{V}) as follows:

\begin{equation}
X_{mC} \equiv \sum_{n=\mu}^{M}\left(\frac{a}{r} \right)^{n}
\bar{C}_{nm} \bar{P}_{nm}(\Theta), 
\qquad X_{mS} \equiv \sum_{n=\mu}^{M}\left(\frac{a}{r} \right)^{n}
\bar{S}_{nm} \bar{P}_{nm}(\Theta), 
\label{xm}
\end{equation}

\begin{equation}
\Omega_{m} \equiv \cos (m\lambda) X_{mC} +  \sin (m\lambda) X_{mS} \ ,
\label{Omega}
\end{equation}

\par\noindent 
where $\mu $ s an integer that depends on $m$. By now substituting
Eqs. (\ref{xm}, \ref{Omega}) into Eq. (\ref{V}) we finally get

\begin{equation}
V = \frac{GM}{r} +  \frac{GM}{r}\sum_{m=0}^{M} \Omega_{m}.
\end{equation}

The above more compact expression for the potential $V$ allows us to
quickly derive the expression of its gradient, which is needed for
solving the body's equation of motion.  We first evaluate the
gradients of the potential with respect to the spherical coordinates
$\lambda$, $\Theta$, and $r$. The gradient with respect to $\lambda$
may be computed by using the following expression
\begin{equation}
V_{\lambda} \equiv \frac{\partial V}{\partial \lambda} = - \frac{GM}{r}
\sum_{m=0}^{M}  m\left[\sin (m\lambda) X_{mC} -  \cos (m\lambda)
  X_{mS}\right] \ ,
\end{equation}
\par\noindent 
with $X_{mC}$ and $X_{mS}$ given by equation \ref{xm}. The computation
of the gradients with respect to $\Theta$ and $r$ require instead the
expressions of the first derivative of the Legendre polynomial,
$\bar{P}^{1}$:

\begin{equation}
X_{mC}^{\theta} \equiv \sum_{n=m}^{M}\left(\frac{a}{r} \right)^{n}
\bar{C}_{nm} \bar{P}_{nm}^{1}(\Theta), 
\qquad 
X_{mS}^{\theta} \equiv \sum_{n=m}^{M}\left(\frac{a}{r} \right)^{n} \bar{S}_{nm} \bar{P}_{nm}^{1}(\Theta)
\end{equation}

\begin{equation}
V_{\theta} \equiv \frac{\partial V}{\partial \theta} = \frac{GM}{r}
\sum_{m=0}^{M} \left[\cos (m\lambda) X_{mC}^{\theta} +
\sin (m\lambda) X_{mS}^{\theta}\right] \ ,
\end{equation}

\begin{equation}
X_{mC}^{r} \equiv \sum_{n=m}^{M}\left(\frac{a}{r} \right)^{n}
(n+1)\bar{C}_{nm} \bar{P}_{nm}(\Theta), 
\quad X_{mS}^{r} \equiv \sum_{n=m}^{M}\left(\frac{a}{r} \right)^{n}
(n+1)\bar{S}_{nm} \bar{P}_{nm}(\Theta) \ ,
\end{equation}

\begin{equation}
V_{r} \equiv \frac{\partial V}{\partial r} = - \frac{GM}{r^{2}}
\left\{1 + \sum_{m=0}^{M} \left[\cos (m\lambda) X_{mC}^{r} +
\sin (m\lambda) X_{mS}^{r}\right] \right\} \ .
\end{equation}

The recursion expressions for the Legendre polynomials and their derivatives are
given in Kuga and Carrara~\cite{kuga}. Finally, the transformation
from spherical to Cartesian coordinates takes into account the partial
derivatives that relate them and it is given by the following formulas:

\begin{eqnarray}
\ddot{x} &=& u\,V_{r}\cos\lambda - \frac{t}{r}\, V_{\theta}\cos \lambda
-  \nonumber V_{\lambda} \frac{\sin \lambda}{u\, r} \\
\ddot{y} &=& u\,V_{r}\sin\lambda - \frac{t}{r}\, V_{\theta}\sin \lambda
+   V_{\lambda} \frac{\cos \lambda}{u\, r}   \\
\ddot{z} &=& t\,V_{r}+\frac{u}{r} V_{\theta} \nonumber
\end{eqnarray}

\par\noindent 
with $t = \cos(\theta)$ and $u = \sin(\theta)$ . The implemented
codes~\cite{kuga} were tested up to order 2159 and degree 2190 without
any noticeable flaw. Although some numerical degradation near the
poles could be expected, we found no significant degradation up to
$\pm 89.999999^{o}$ latitude. Also sample cases were used as tests
for accuracy, performance and reliability of our geopotential
algorithm for integrating Earth orbits.

The other major gravitational perturbations taken into account were
those due to the Sun and the Moon, whose gravitational potentials can
be represented by point-mass models. The acceleration due to a
point-mass is modeled by the usual Newtonian expression
\begin{equation}
\ddot{\bf r}_{Sun, Moon} = \mu_p \ \left(\frac{{\bf r}_p - {\bf
      r}}{|{\bf r}_p - {\bf r}|^3} - \frac{{\bf r}_p}{|{\bf r}_p|^3} \right) \ ,
\label{EM}
\end{equation}
where $\mu_p$ is the gravitational coefficient ($G M_p$) of the
perturbing body, and ${\bf r}_p$ is the inertial position vector of
the perturbing body. Note that the inertial coordinates of the Sun and
the Moon are obtained analytically, and are characterized by an
accuracy of $10^{-3}$ degrees for the Sun and $10^{-2}$ degrees for
the Moon.  For the numerical integration of the orbit the
predictor-corrector algorithm ODE \cite{Shampine} with variable order
and step-size was used to tolerances set around $10^{-12}$. The set of
first order differential equations for position, ${\bf r}$, and
velocity, ${\bf v}$, were integrated in the J2000 inertial coordinate
system, and they are given by the following expressions
\begin{eqnarray}
\dot{\bf r} &=& {\bf v} \ ,
\nonumber
\\
\dot{\bf v} &=& {\bf a}_g + {\bf a}_{Sun} + {\bf a}_{Moon} \ ,
\nonumber
\end{eqnarray}
where ${\bf a}_g$ is the geopotential acceleration, an d ${\bf
  a}_{Sun}$, ${\bf a}_{Moon}$ are the corresponding perturbing
gravitational accelerations due to the Sun and Moon
respectively. Where needed, care was taken to compute the
transformation between the J2000 inertial system and ITRF system using
the SOFA library \cite{SOFA}.

\end{document}